\documentclass[10pt]{bmc_article}    
\usepackage{multirow}
\usepackage{float}
\usepackage{graphicx}
\usepackage{wasysym}
\usepackage{amssymb}
\usepackage{lscape}
\usepackage{stmaryrd}

\usepackage{color,soul}
\usepackage{url}
\usepackage[figuresright]{rotating}
\usepackage{cite} 
\usepackage{url}  

\newenvironment{bmcformat}{\baselineskip20pt\sloppy\setboolean{publ}{false}}{\baselineskip20pt\sloppy}

\newboolean{publ}

\begin{document}
\begin{bmcformat}
	
\title{Mapping structural diversity in networks sharing a given degree distribution and global clustering: Adaptive resolution grid search evolution with Diophantine equation-based mutations}

\author{
    Peter Overbury
    \and
    Istv\'an Kiss
    \and 
    Luc Berthouze\correspondingauthor
    \email{Luc Berthouze\correspondingauthor - L.Berthouze@sussex.ac.uk}
}

\address{
Department of Informatics, University of Sussex, UK
}

\maketitle

\begin{abstract}

Methods that generate networks sharing a given degree distribution and global clustering can induce changes in structural properties other than that controlled for. Diversity in structural properties, in turn, can affect the outcomes of dynamical processes operating on those networks. Since exhaustive sampling is not possible, we propose a novel evolutionary framework for mapping this structural diversity. The three main features of this framework are: (a) subgraph-based encoding of networks, (b) exact mutations based on solving systems of Diophantine equations, and (c) heuristic diversity-driven mechanism to drive resolution changes in the MapElite algorithm. We show that our framework can elicit networks with diversity in their higher-order structure and that this diversity affects the behaviour of the complex contagion model. Through a comparison with state of the art clustered network generation methods, we demonstrate that our approach can uncover a comparably diverse range of networks without needing computationally unfeasible mixing times. Further, we suggest that the subgraph-based encoding provides greater confidence in the diversity of higher-order network structure for low numbers of samples and is the basis for explaining our results with complex contagion model. We believe that this framework could be applied to other complex landscapes that cannot be practically mapped via exhaustive sampling.

\end{abstract}

\section{Introduction}

Almost any complex system involving the interaction of constituent components can be represented as a network and networks have become a paradigm of choice for modelling and analysing such systems. It is now well known that node-level and structural properties of networks (e.g., degree-distribution, assortativity, clustering,  modularity) can fundamentally affect the way the system operates~\cite{newman2002spread,read2003disease,pastor2015epidemic,kiss2017mathematics}. Clustering, in particular, has been the subject of much work with both empirical and analytical results~\cite{watts1998collective,eames2008modelling,green2010large}. 
However, there is also some growing awareness that local structure (e.g., subgraph composition) may also have an important impact on dynamics~\cite{house2009motif,karrer2010random,Ritchie2017GenerationMatters}. 

To be able to show this more compellingly, there is a need for methods to sample the space of networks satisfying set constraints, e.g., degree sequence, assortativity, global clustering. Currently available network generative methods can be categorised in terms of where they fall within the `one-shot' - `growing/developmental' spectrum (see~\cite{Betzel20170623} for a more comprehensive treatment). On the `one-shot' end of the spectrum, an algorithm produces a single network. One of the most popular example of such generative models is the exponential random graphs model~\cite{robins2007introduction} whereby one assumes that links are random variables and each realisation comes from a probability distribution of graphs of a given number of nodes. 
Variants of such models that provide more control over relationships between nodes include simplicial network models~\cite{courtney2016generalized} and graphons~\cite{lovasz2012large}.  
At the other end of the spectrum, one can find methods that fundamentally involve rewiring existing networks and are typically based on Markov Chain Monte Carlo processes~\cite{DelGenio2010EfficientSequence}. To construct networks with fixed degree sequence and global clustering coefficient, BigV~\cite{house2010impact} starts from a random network and performs a series of degree-preserving rewiring operations which increase clustering. The process is repeated until the desired clustering coefficient is achieved.  
Conversely, dk-series decomposition~\cite{orsini2015quantifying} uses rewiring to generate randomised versions of a given network that preserve network characteristics from average degree (k=0) to global clustering coefficient (k=2.1). In principle, rewiring approaches could be used to sample the network space, however, they do not actually provide any control over which local higher-order structure property is being changed. Further the question remains of whether these approaches will necessarily cover the full range of possible networks depending on the seed network~\cite{orsini2015quantifying}. Two common features to both approaches are that (a) there is great computational cost to mapping the network space and (b) they do not lend themselves well to controlling / assessing the make-up of networks beyond the specified characteristics.

An alternative approach is therefore to not attempt to being as exhaustive as possible but rather to maximise the diversity of networks found within a given amount of time. This can be couched in terms of a multi-objective optimisation problem. In the kind of scenario we consider, population-based algorithms (see ~\cite{giagkiozis2015overview, Zhou2011MultiobjectiveArt} for reviews, and also~\cite{liu2017investigating}) can prove particularly helpful. More specifically, there has been a growing body of research into so-called quality diversity (or illuminative) algorithms, whose focus is to discover both quality and diversity at the same time, see ~\cite{pugh2016quality,Cully2017QualityFramework} for example. In the MapElite approach~\cite{Mouret2015IlluminatingElites}, the space of features (or behaviours) an individual might possess is divided into cells that act as niches to the population, forcing new individuals to only compete with individuals in the same cell. As a result, only the fittest (the elite) in each cell remain in the population, thus providing a collection of diverse, high-performing individuals. In previous work~\cite{Overbury2017ComplexV}, we combined the MapElite method with CMA, the cardinality matching algorithm~\cite{Ritchie2017GenerationMatters}. CMA breaks the problem of generating networks down to the subgraph level (subgraphs being small structures) arranging set populations of these subgraphs in such a way to satisfy a prescribed degree distribution and global clustering coefficient. Although our method elicited a wide range of diversity at both structural and behaviour levels compared with other methods of network generation, it suffered from being very slow and ineffective at producing large pools of networks. 

In this paper, we substantially improve the method through two major changes:
\begin{enumerate}
	\item an exact Diophantine equation-based \textit{mutation} method that guarantees that all individuals in the population are fit, i.e., they satisfy the constraints,  
	\item an adaptive resolution mechanism whereby the size of a niche changes during evolution in response to the level of variation between individuals. This allows us to efficiently control the trade-off between coverage and diversity.  
\end{enumerate}

In what follows, we detail both mechanisms after a brief reminder of how networks are encoded.

\section{Methods}\label{sec::methods}
\subsection{Defining the search space: Network encoding\label{ssec:encoding}}
Manipulating diversity in higher-order structure whilst maintaining degree sequence and global clustering coefficient requires a parametrisation of networks (and a generative mechanism) that is both parsimonious enough that it enables systematic exploration and complete enough that it allows explicit control of global features. As in~\cite{Overbury2017ComplexV}, we encode networks in terms of the population counts of each subgraph in an arbitrarily chosen family of subgraphs, e.g., $\{\triangle, \boxempty, \boxslash, \boxtimes, \pentagon\}$, provided this family contains at least one clustering-inducing subgraph. To generate networks from a given sequence of subgraph counts, we use the cardinality-matching algorithm (CMA)~\cite{Ritchie2017GenerationMatters}. 
It is important to stress at the outset that the algorithm is not exact: first, to mitigate the the combinatorial complexity of satisfying all constraints, it is necessary to specify a fraction of edges not accounted for by the subgraphs (free edges); second, as described in~\cite{Ritchie2017GenerationMatters}, the allocation process (particularly when free edges are involved) can lead to by-products. For example, the addition of a free edge can lead to two distinct $\triangle$ turning into one $\boxslash$ and one $\triangle$. Figure~\ref{fig:fig1_2} illustrates this problem by showing that whilst CMA yields fairly good control over the $\boxtimes$, there is more noise for $\triangle$ and $\boxslash$ (note that by-products of non clustering-inducing subgraphs is not an issue since they do not have any impact over the clustering coefficient). Nevertheless, the right-hand side panel in Figure~\ref{fig:fig1_2} demonstrates that despite the by-products, the process yields acceptable control over the global clustering coefficient with the obtained clustering values never exceeding the target value by more than 0.003, i.e., at most 21 $\triangle$ in a regular network of size N=1000 and degree k=7. 

\begin{figure}
\includegraphics[width=\textwidth]{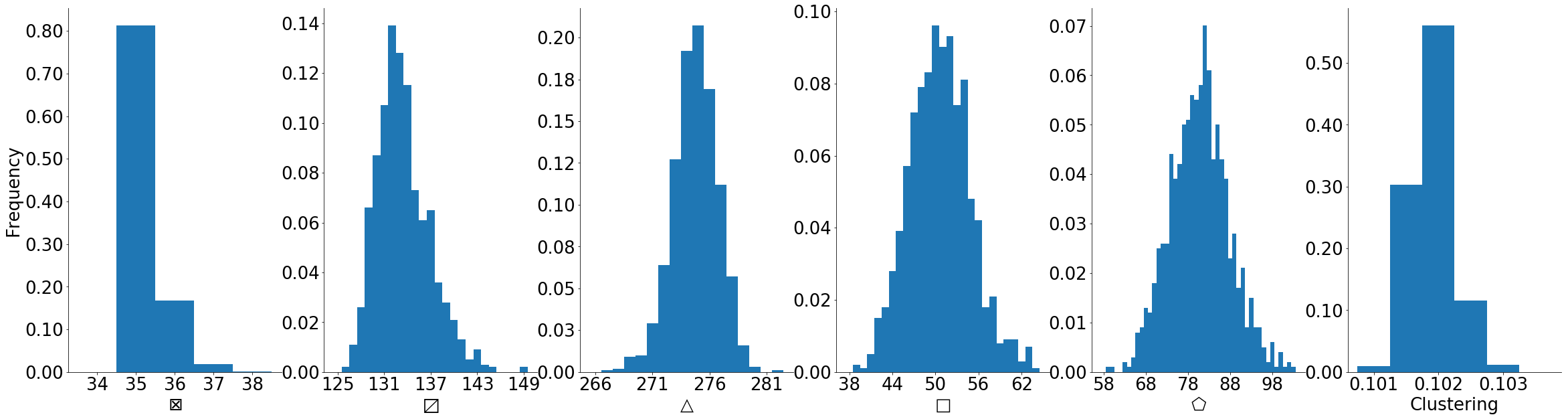}
\caption{First five panels: Histograms of the prevalence of each subgraph for 1000 networks generated by CMA for specification ($\boxtimes$:35, $\boxslash$:128, $\triangle$:277, $\boxempty$:35, $\pentagon$:42). The subgraphs were counted using the method in~\cite{Ritchie2014Higher-orderNetworks}. The number of $\triangle$ denotes the number of $\triangle$ not involved in any other clustering-inducing subgraphs. Last panel: Histogram of global clustering coefficient values~\cite{Keeling2011NetworksDisease} for the 1,000 networks considered. The target global clustering coefficient was 0.1.} \label{fig:fig1_2}
\end{figure}

\subsection{Defining movement within the search space: Exact mutations\label{ssec:dioph}}
In a standard evolutionary framework, mutations involve random changes in the make-up of each individual. Here, such approach is extremely wasteful because preserving the global properties of the network impose constraints on what changes are possible in one or more subgraphs given a change in another. For example, a constant global clustering coefficient imposes that the addition of one $\boxslash$ come with the loss of two $\triangle$. However, simply reducing the number of $\triangle$ by 2 does not suffice because such operation would leave a deficit of 1 edge at network level. In this paper, we cast the problem of identifying degree- and clustering-preserving mutations (exact mutations, thereafter) in terms of solving a Diophantine problem, i.e., finding the integer solutions to an undetermined system of linear equations. Formally, an exact mutation is an integer solution of the system $A\mathbf{x}=b$ where: $\mathbf{x}$ is a column vector of $n$ rows and specifies the change in the number of each of the subgraphs specifying the network (i.e., n is the cardinality of the family of subgraphs used to parameterise the networks; n=5 throughout the paper); $A$ and $b$ are a $3 \times n$ matrix and a column vector of 3 rows, specifying the 3 constraints that a mutation $\mathbf{x}$ must satisfy, namely: (i) the change in the total number of triangles in the network must be 0, (ii) the change in the total number of edges in the network must be 0, (iii) the size of the change for the subgraph count(s) being mutated has the required size (see below). Note that the third constraint is purely for programming convenience as only the first two rows specify constraints between subgraphs. To illustrate the principle, given individual ($\boxtimes$:61, $\triangle$:283, $\boxempty$:110, $\pentagon$:142, $\boxslash$:87) and a required mutation of size 2 in the number of $\boxslash$, a possible vector $\mathbf{x}$ is ($\boxtimes$:-1,$\triangle$:0,$\boxempty$:-1,$\pentagon$:0,$\boxslash$:2) leading to the new network specification ($\boxtimes$:60, $\triangle$:283, $\boxempty$:109, $\pentagon$:142, $\boxslash$:89). It is easily verified that the gain of four $\triangle$ via the addition of two $\boxslash$ is compensated by the loss of one $\boxtimes$, whereas the resulting excess of 4 edges ($2\times 5-1\times 6$) is absorbed by the loss of one $\boxempty$. 

Solving an underdetermined system of Diophantine equations in general is a hard problem, however, finding solutions with the lowest Euclidean norm is easier~\cite{havas1998extended}. We used the following implementation: \url{http://github.com/tclose/Diophantine}. There are two implications to this. The first is that it makes an exhaustive search impossible (although this is not strictly the aim of this work). The second is that solutions tend to be homogeneous (with little difference between absolute value of the components of the solution), which significantly biases how the space of solutions is sampled. For this reason, we built a catalog of solutions by systematically enforcing values for each component of the solution vector. In the experiments that follow, the catalog of possible mutations for 7-regular networks with global clustering of 0.1 comprised 1582 exact mutations with mutation sizes ranging from 1 to 128 and involving between 1 and 3 subgraphs.  

There are two additional observations to be made. First, because all computations are integer, given a particular family of subgraphs, some mutations are not possible (i.e., the solver returns no solutions). A trivial example is that given a family where the only clustering-inducing subgraphs are $\triangle$, $\boxtimes$ and $\boxslash$, it is not possible to mutate the number of triangles by an odd number. Second, even if there is a solution, there is no guarantee that the network thus specified will be graphical~\cite{erdos1960grafok} or realisable (in a configuration model sense). In our implementation, we leave it to the CMA algorithm to make this determination. 

\subsection{Adaptive resolution change mechanism}\label{ssec:adap_res}
A challenge with MapElite and indeed any novelty-driven method of search that involves cells / niches is the selection of a suitable cell / niche size in the absence of prior knowledge about the space. Here, a particular feature of the problem at hand is that there is a trade-off between maximising novelty by discovering as many (valid) network descriptions as possible and maximising novelty through exploring the diversity existing in network realisations satisfying a single network specification (subgraph decomposition). To do so, we propose a novel adaptive resolution mechanism defined as follows: 
\begin{itemize}
	\item Start search at the same, low, resolution across all dimensions (where the number of dimensions is the cardinality of the family of subgraphs used to parameterise networks). 
	\item When the ratio between the number of cells being revisited (by mutations) and the number of new cells being discovered exceeds a threshold (set to 2 in our experiments), halve the resolution (across all dimensions) of a number of the cells with the highest measure of interestingness (see Section~\ref{ssec:interestingness}), and adjust mutation size (for those cells) accordingly.  
\end{itemize}
Two critical components of this mechanism are the measure of interestingness and the relationship between mutation size and cell resolution. They are detailed below. 

\subsubsection{Measure of interestingness}\label{ssec:interestingness}
Changing the resolution of a cell means increasing the likelihood of exploring this area of the space and therefore a criterion is needed that reflects the value of this cell in maximising the second component of the novelty described above, namely, diversity in the structure of network realisations of a single network specification. We propose for this criterion to be the variance in a measure of network structure that is not uniquely determined by the network' subgraph decomposition. In our experiments we used betweenness centrality~\cite{freeman1977set} although others could be used equally. 

Practically, each cell maintains a copy of the specification of the fittest individual (since with varying resolution, the cell only specifies a range of values for each dimension of the specification), along with the variance in the measure of interestingness calculated over all individuals sampled when the cell was visited. 

\subsubsection{Relationship between mutation size and cell resolution}\label{ssec:size_mut_cell}
The value of adaptive mutation mechanisms in controlling the trade-off between exploration and exploitation is well established~\cite{eiben1999parameter}, including within the MapElite framework (see~\cite{nordmoen2018dynamic} for example). Here, we link the range within which a mutation size is selected to the resolution of the cell in which an individual exists; specifically between 1 (minimum) and 2 (maximum) cell sizes. This guarantees that mutations are small enough to preserve locality (excessively large mutations would lose the benefit of locally heterogeneous resolution) whilst ensuring that any mutation will result in a different cell being explored (since at least one dimension of the network specification will change by at least one cell). 

As time increases, the total number of cells in the space will increase such that the average cell size will decrease, and with it, the average mutation size. This means that the evolution process gradually moves from a global search to a local search with emphasis on those areas of the space yielding most diversity. 

\subsubsection{Implementation}
Although the idea of starting with a coarse discretization and then increasing granularity was mentioned by the authors of the MapElite framework ~\cite{Mouret2015IlluminatingElites}, we are not aware of any such implementation and further we are not aware of any discussion as to the computational requirements. Indeed, even in those papers in which cell size is a point of interest, e.g. ~\cite{vassiliades2017using,vassiliades2017comparison}, the total number of cells is known a priori. For a mechanism such as ours to be computationally tractable in a high-dimensional search space, there is a need for efficient operations for adding, deleting and updating cells. Our implementation relies on a tree data structure developed in-house and available at \url{https://github.com/harrygcollins/TreeBasedGA}. 
\section{Results}\label{sec::results}
To allow comparison with previous work, all results that follow concern the exploration of homogeneous networks with $N=1000$, degree $k=7$ and global clustering coefficient $C=0.1$. All runs started from the same starting population of 5 (randomly picked) valid CMA-generated networks. 
We analysed the impact of our methodological changes in terms of 3 measures: 
\begin{itemize}
\item rate of discovery: the number of iterations needed to get a number of networks, 
\item quality of discovery: the diversity in network specifications uncovered,
\item behavioural diversity: whether network diversity impacts dynamics.
\end{itemize}

\subsection{Impact of Diophantine-based mutation on rate of discovery}
\begin{figure}
\centering\includegraphics[width=0.7\textwidth]{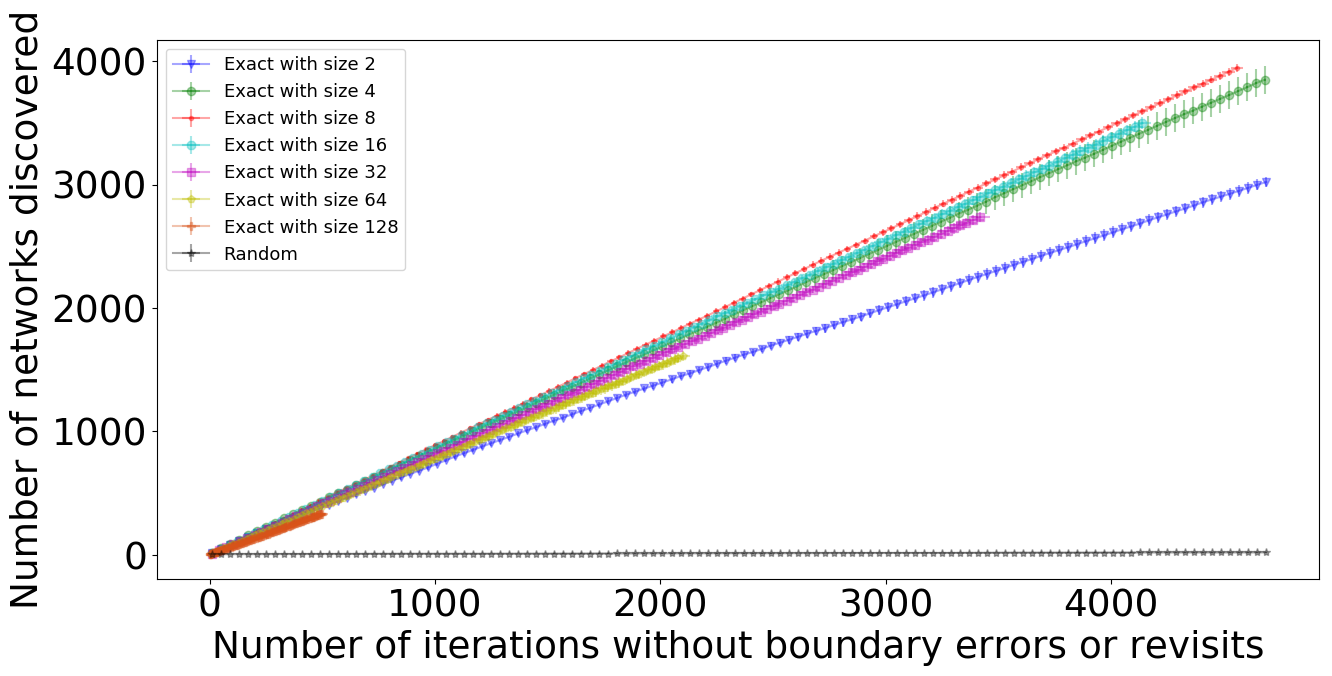}
\caption{Rate of discovery of valid networks when exact (with different mutation sizes) and random mutations are used.}
\label{fig:fig2}
\end{figure}

To characterise the impact of the use of exact mutations on the search process, we compare it with the baseline of random mutations. Setting the cell resolution to its maximum (i.e., one cell per network specification), we systematically varied mutation size between 2 and 128 in powers of 2. For random mutations, the size was randomly picked.   
In all cases, we evaluated the rate at which new (valid) networks were discovered as a function of mutations (4700 in all cases) as well as the diversity in network specifications (the coverage). As shown by Figure~\ref{fig:fig2}, there is a significant gain in speed (and number of networks obtained) when using exact mutations, irrespective of the mutation size. 
It is worth noting that a higher rate of discovery does not necessarily result in a greater number of networks. This is because a substantial number of iterations are lost, either due to out-of-bounds mutations or higher rate of revisits.

Figure~\ref{fig:fig3} shows the frequencies at which subgraphs occur for both exact and random mutations. Whilst random mutations show fairly uniform frequencies, coverage of the space is extremely patchy due to the difficulty obtaining realisable networks. In contrast, exact mutations lead to dense coverage of the space (including beyond that sampled by the random mutations). The Figure clearly shows the impact of mutation size with small mutations (e.g., size 2) resulting in well defined peaks of higher frequency and large mutations (e.g., size 128) resulting in a more uniform histogram although the number of networks found drops significantly (as explained before). There is therefore a balance to be reached which will be explored below via the adaptive resolution mechanism. In what follows, all experiments start with a cell size of 64 (as a cell size of 128 leads to too few networks). 

\begin{figure}[t]
\includegraphics[width=1\textwidth]{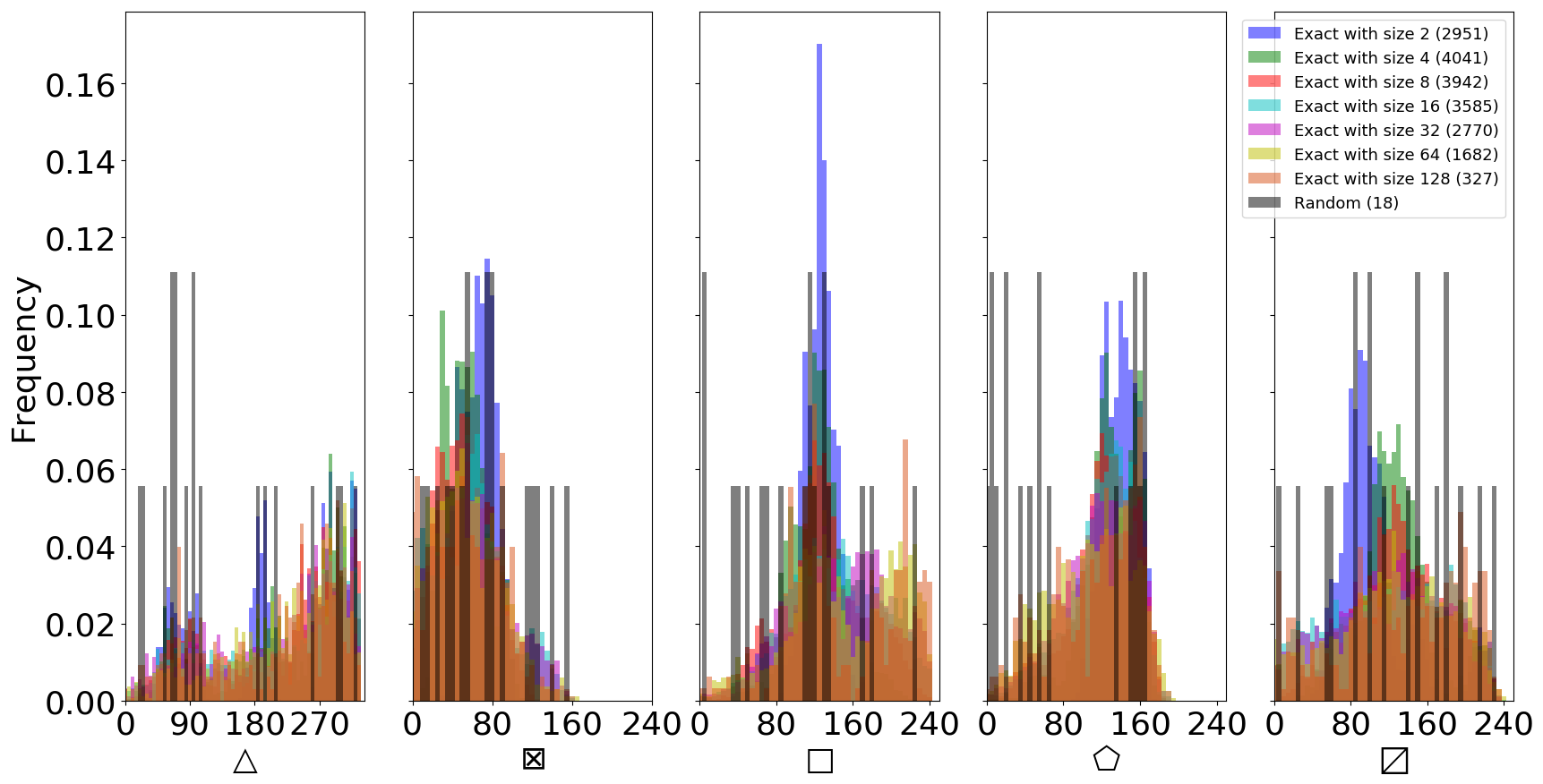}
\caption{Frequencies of subgraph population counts for both exact (with mutation sizes from 2 to 128) and random mutations. Bin size was set to 5 for all subgraphs. The number of networks found by each method after 4700 iterations is provided in the legend.}
\label{fig:fig3}
\end{figure}

\subsection{Impact of adaptive resolution search on quality of discovery}\label{ssec::adapt_res_results}

To illustrate the benefit of using adaptive rather than fixed resolution search, we compared the network specifications discovered by our method with those obtained using either a fixed mutation size of 64 (coarsest resolution enabling greatest coverage) or a fixed mutation size of 8 (shown previously to yield the highest rate of discovery). As shown by Figure~\ref{fig:fig4}, networks uncovered using the adaptive resolution search showed the largest breadth of subgraph counts, e.g., largest range of $\boxtimes$. Interestingly, even though using a fixed mutation size of 8 yields a much larger ensemble of networks (almost 20 times larger than using either our method or fixed resolution size of 64), we observe fairly unimodal distributions of subgraph counts reflecting the lack of coverage. 

\begin{figure}
\includegraphics[width=\textwidth]{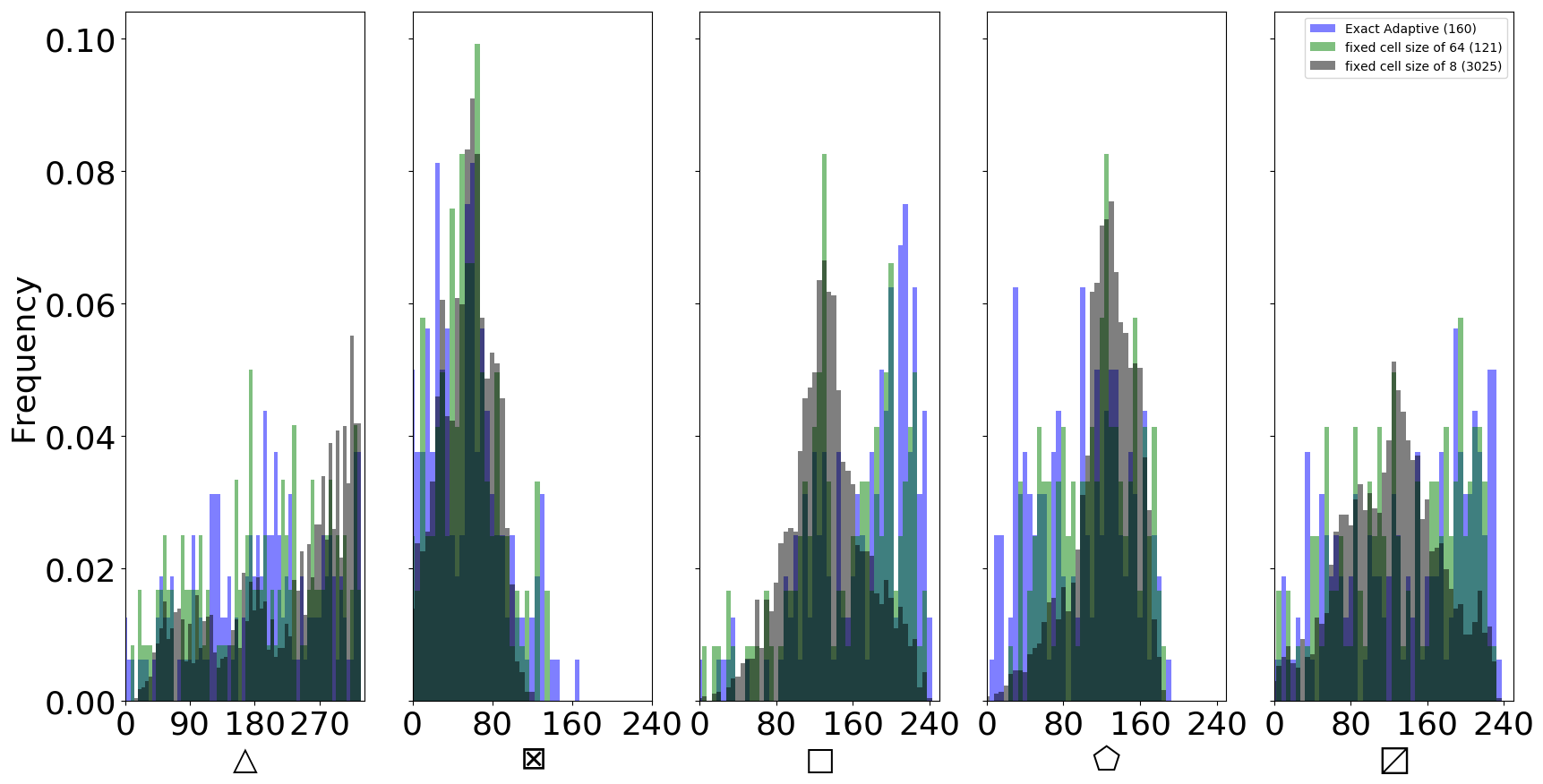}
\caption{Frequencies of subgraph population counts for proposed method (blue), fixed mutation size of 64 (green) and fixed mutation size of 8 (black). Bin size was set to 5 for all subgraphs. The number of networks found by each method after 4700 iterations is provided in the legend.}
\label{fig:fig4}
\end{figure}

To address the concern that such differences may be a random artefact, we assessed the range of betweenness centrality found in the above networks and that of an identical number of networks generated using BigV and dk-2.1 randomisation. Both methods were used so as to maintain the same distribution of global clustering coefficients (see right panel in Figure~\ref{fig:fig5}), i.e., the BigV rewiring process was stopped when the required clustering coefficient was reached and the dk2.1 randomisation was seeded by CMA-generated networks with the required clustering coefficient. Nevertheless, as shown by Figure~\ref{fig:fig5} (left panel), we found the networks to span a different range of betweenness centrality values. An explanation for this finding will be provided elsewhere (it pertains to CMA seeking to prevent subgraphs around a node from sharing edges) but this demonstrates that the networks are structurally different. To our knowledge, this is also the first evidence that, as considered by its authors, dk2.1 randomisation may not provide uniform sampling. 

\begin{figure*}[t]
\includegraphics[width=1\textwidth]{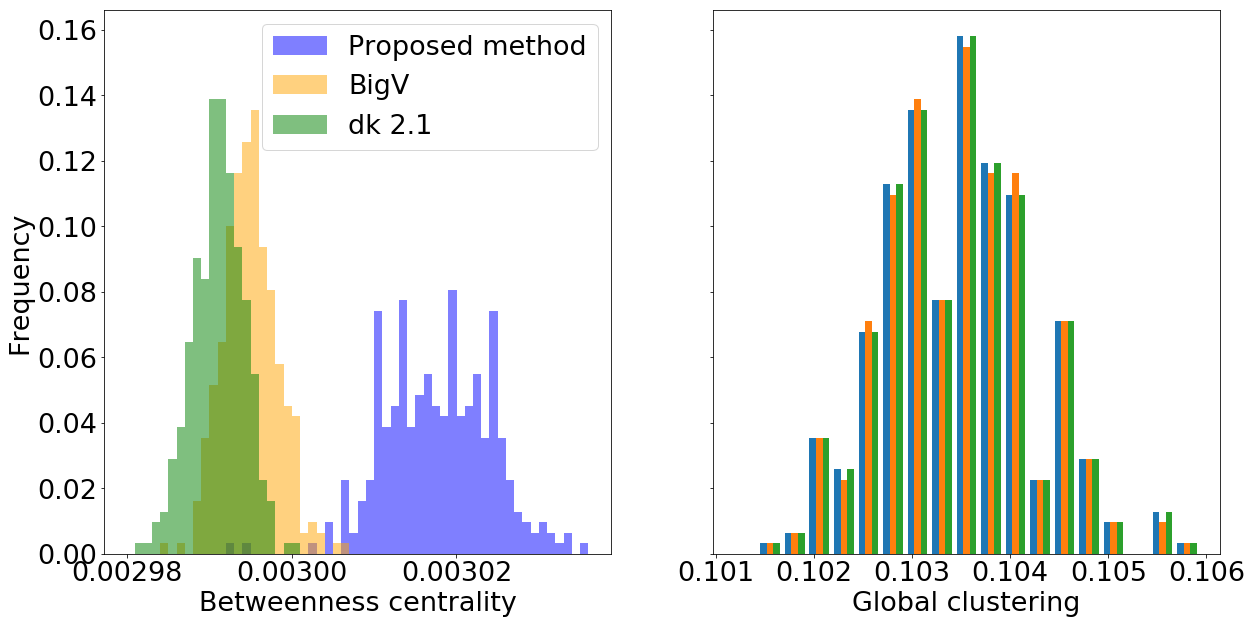}
\caption{Histograms of betweenness centrality (left) and clustering coefficients (right) for GA, BigV and dK-generated 7-regular networks.} \label{fig:fig5}
\end{figure*}

\subsection{Impact of diversity on behaviour: Complex contagion}
\label{ssec::beheffect}
To illustrate that diversity in such higher-order structure does impact behaviour, we consider the complex contagion model~\cite{miller2015complex}. This model differs from a classical SIR (susceptible-infected-recovered) epidemic by requiring that susceptible nodes are exposed to multiple infectious events before becoming infected. Further, these events must be from different infectious neighbours as only the first infection attempt from an infectious node counts; and infected individuals remain infected for the duration of the epidemic. This dynamics is known to exhibit a critical transition in relation to the number of infected nodes at the start of the epidemic. In preliminary work, we showed that given a degree distribution and a global clustering coefficient, the parameter value at which the transition occurred could fluctuate. 

Here, we compared the range of parameter values at which the transition occurred for maximally different (Euclidean distance in their subgraph counts) pairs of networks found (a) through random exploration and (b) through our proposed search mechanism. For each of the networks and for each parameter value, we ran 100 simulations to robustly identify the critical transition. Figure~\ref{fig:fig6} shows a substantially greater range of parameter values for networks found through the proposed search mechanism, thus confirming greater diversity was achieved with our method (even though the computational cost of eliciting the same number of networks through random mutation was far greater).  
 
\begin{figure}[t]
  \includegraphics[width=1\linewidth]{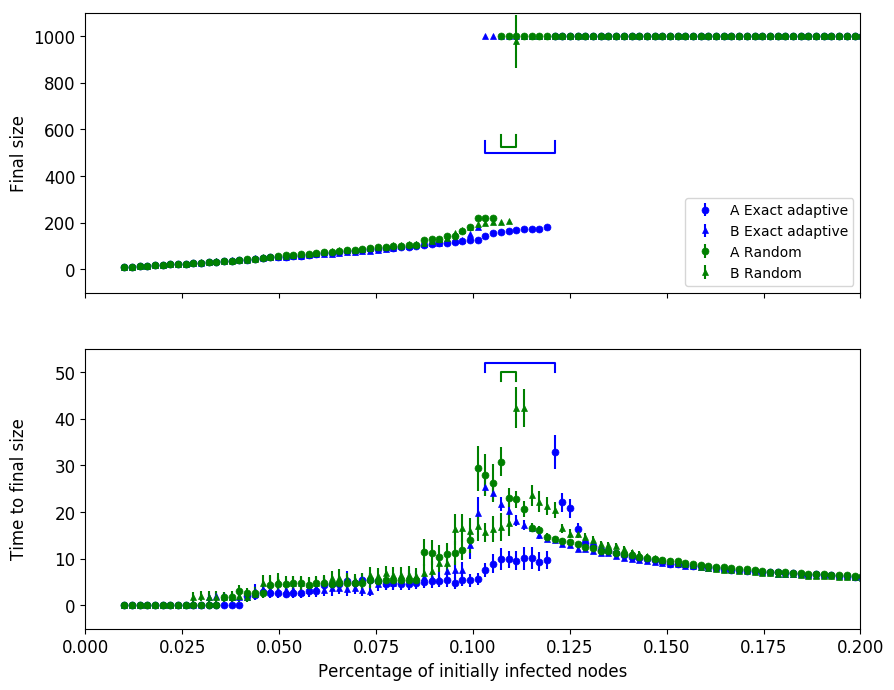}
  \caption{Ranges of parameter values over which the critical transition of complex contagion simulations occurs in networks found by the proposed search mechanism (blue) and random mutations (green). The critical transition is identified as the value parameter at which there is maximal variability in both final size and the time needed to reach this final size. All $N=1000$ individuals had threshold $r=3$ and transmission rate $\beta=1.0$.}
  \label{fig:fig6}
\end{figure}

\section{Discussion}
In this paper, we presented a computational framework that makes it possible to explore structural diversity in networks sharing a set of properties. Although we only provided examples when specifying degree distribution and global clustering coefficient, the framework is applicable to other scenarios.  Differently from classical network generating methods, which rely on very long mixing times to provide uniform coverage, our approach borrows concepts from evolutionary algorithms to more rapidly identify interesting regions of the solution space, i.e., regions of the space containing structurally more diverse networks. Encoding networks based on their subgraph decomposition provides control over the local structure around nodes. Our experiments have revealed that our methodology makes it easier to elicit structural differences that have an impact on dynamics running on the networks. A promising line of enquiry is to systematically study the importance of a given subgraph on dynamics by restricting the search process to mutations that increase/decrease the number of this subgraph. Another line of further work involves exploring other measures of interestingness, e.g., local clustering diversity, vertex-level entropy. Finally, proper sensitivity analysis on a number of parameters is required but beyond the scope of this paper.

{\ifthenelse{\boolean{publ}}{\footnotesize}{\small}
\bibliographystyle{bmc_article}
\bibliography{Mendeley} 
}

\end{bmcformat}
\end{document}